# Neuroinformatics:
# What are us, where are we going, how to measure our way?

A.N. Gorban'

Institute of Computational Modelling of SB RAS
Krasnoyarsk-36, 660036, Russia
E-mail: gorban@cc.krascience.rssi.ru

What is neuroinformatics? For me here and now neuroinformatics is a direction of science and information technology, dealing with development and study of the methods for solution of problems by means of neural networks. A base example of artificial neural network, which will be referred to below, is a feed-forward network from standard neurons (fig. 1)

A field of science cannot be determined only by fixing what it is "dealing with". The main component, actually constituting a scientific direction, is "THE GREAT PROBLEM", around which the efforts are concentrated. One may state even categorically: if there is no a great problem, there is no a field of science, but only more or less skilful imitation (the question, who and what for needs imitations, is an amusing one, but it can lead us far away). Just main problems impart sense to each particular study.

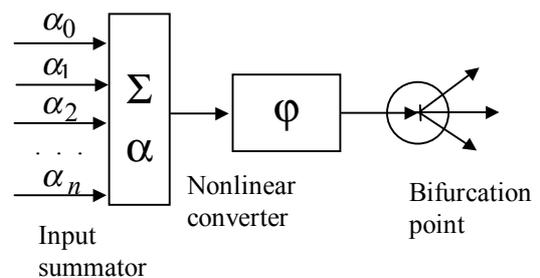

*Fig. 1. Standard neuron.*

What is "THE GREAT PROBLEM" for neuroinformatics? What imparts sense to this vast flow of studies? The answer will be as we state it – it is impossible to separate a reflexive exploration of sense of an activity from generation and ascription of the sense "post factum".

There are two directions of search for the sense – first, study of brain (solution of mysteries of thinking), and, second, problems of efficiency of computations. Many people are inspired by the banner of "brain-like" computers, but for me it is alien banner. I lack the determinacy of problem here – the determinacy of gap in activity: what we cannot do now and what we will be able to do after obtaining the solution? Brain-like computer is rather a metaphor than designation of the problem. However, each has its own symbol of belief, though those are not numerous.

*The problem of effective parallelism* pretends to be the central problem which is being solved by the whole neuroinformatics. Long ago sank into oblivion the naive idea: let take more processors, and the efficiency of computer will increase proportionally. There is a well known "Minsky hypothesis" (fig. 2): efficiency of a parallel system increases (approximately) proportionally to logarithm of the number of processors; at least, it is a convex upwards (i.e. concave) function (discussion of these problems can be found in [2]).

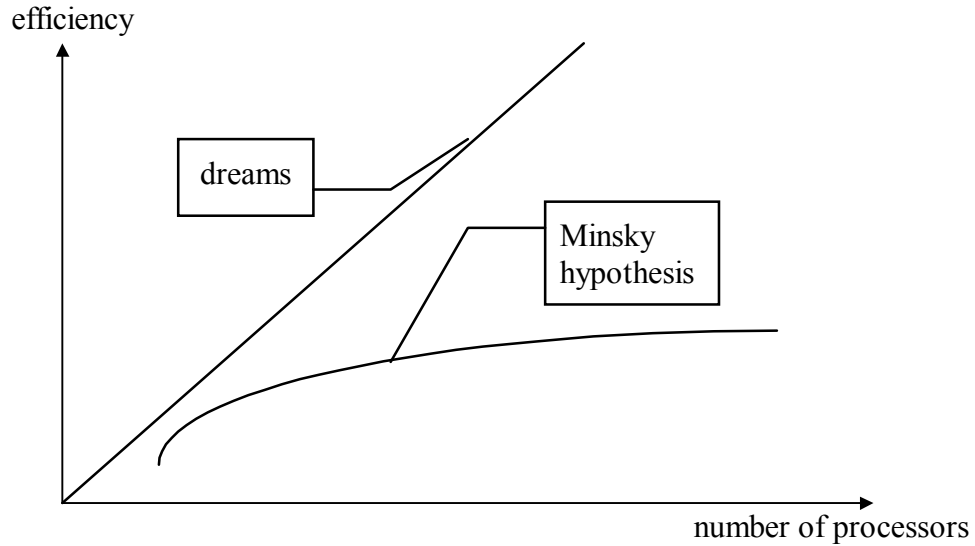

*Fig. 2. Minsky hypothesis*

Nowadays more and more often as a way to overcome this limitation the following approach is used: for different classes of problems maximum parallel algorithms of solution are constructed which use some abstract architecture (paradigm) of fine grain parallelism, and for concrete parallel computers means are developed for realization of parallel processes of a given abstract architecture. As a result, an efficient technique of production of parallel programs is created. Neuroinformatics supplies universal fine grain parallel architectures for solution of different classes of problems. Thus, it is used for solving the problem of efficient parallelism according to the following scheme (fig. 3).

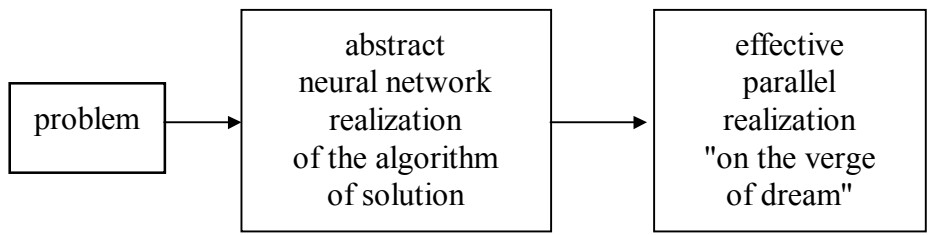

*Fig. 3. Way to effective parallelism.*

One of main problems of brain studies and creation of "brain-like" devices is the same problem of effective parallelism: how the brain uses its billion of neurons for effective parallel operation? So, contradictions between the different ways of search for sense are not so large.

In addition to "the great problem" of effective parallelism, let suggest a great applied problem, which can be solved on this way. **Controlled thermonuclear synthesis** demands for new tools for retaining hot plasma, and, in general, for controlling its state. It seems quite probable that the possibilities of physical traps are exhausted, and any essential increase of stability is unreachable on this way. May be, time has come for creation of a cybernetic system which recognizes instabilities and effectively suppresses them. There exist neuro-equilibrist – a cart driven by self-training neural network, which supports a thin rod in unstable vertical position. Can neural network recognize changes in the state of plasma and rise of instabilities, elaborate and realize the control action over time of the order of $10^{-8}$ s? It is very likely that this is possible. The time of operation of one layer of network can be of the order of $0,3\times10^{-8}$ s, and three – five layers can turn out to be quite sufficient. Then such quick-action is attainable, if there are no extra devices between the network and plasma which cause considerable delay, the information is perceived by the network directly through a system of sensors connected to it, and the signals of output neurons act upon the plasma directly (most probably, through electromagnetic field) – as in the magic "thinking is action". The number of neurons can be estimated as the number of unstable degrees of freedom which must be suppressed. On the whole, such a project seems to be quite realizable by contemporary tools, and not too expensive in comparison with other ways of retaining plasma.

Yet, why we speak only about great problems? Generally, not everything useful and intellectual is a science. For instance, most important in personal computers is neither scientific character nor progress in solution of great problems, but a system of services which they propose to the user. What if to consider neuroinformatics as *a services sector*, what that will result in?

In the system of services rendered by neuroinformatics to actual customer the main place now belongs to data processing, including adaptive signal processing [3]. The central place at the services market is occupied by financial applications [4]. Further are situated military-industrial (and here one cannon find exhaustive review of literature because of understandable reasons), and then medical applications (some examples are given in [5]). In my opinion, the most important achievement of neuroinformatics over the last decade in the sector of intellectual services is systematic use of neural networks *for extraction of knowledge from data*. In the second place I would put the large-scale circulation of *neural net methods of data visualization*, based on self-organizing Kohonen maps [6]. As to *pattern recognition and nonlinear regression problems* of various kinds, nowadays exist tens (or even hundreds) neural network program products which give such opportunity to user, and there are thousands of works in which this opportunity was used.

Neural network extraction of knowledge from data is based on two circumstances: firstly, neural network usually is very redundant – there are too many connections for solving the given concrete problem, and it is possible rarefy the network, constantly training it, so that only the necessary connections would remain. Secondly, rarefied network can be read as explicit algorithm of solution of the problem (for instance, in the form of tree of solutions), – and this is, in essence, explicit knowledge. That is especially simple if one follows certain rules of rarefication ensuring easiness of reading. For instance, at first one can strive for minimization of the number of input connections of all neurons, then decrease the number of neurons and the number of entries of network, and finally, do away with as many connections as possible. Such simplification is done either on the basis of analysis of indices of sensitivity or due to introduction into the evaluation of the quality of training a penalty for complexity of structure. Both these methods were introduced and studied in monograph [7], but other variants of algorithms [8,9][1] are widely spread now.

During training artificial neural network creates some internal rules, but these rules are hidden in the structure of a network and not clear to the user. Moreover, these rules are so difficult for interpretation and understanding. Also discovery of explicit rules for solving of a problem is often more significant than the solution itself. Thus, there is a problem of refining of the hidden knowledge and translating them to natural language.

---

[1] Now as before I wonder why in the algorithms of rarefication widely spread in the West the indices of sensitivity are used which are based on analysis of the second derivatives, whereas it would be much faster and cheaper to use average values of modules of the first derivatives in the course of training [7]. The other used methods of "zero order" which consist in pruning of connections with small weights are less efficient.

The main ideas of generalized technology of extraction of explicit knowledge from data are:
1) maximal reduction of network complexity (not only removal of neurons or synapses, but removal all the unnecessary elements and signals and reduction of the complexity of elements),
2) using of adjustable and flexible pruning process (the pruning sequence shouldn't be predetermined - the user should have a possibility to prune network on his own way in order to achieve a desired network structure for the purpose of extraction of rules of desired type and form),
3) extraction of rules not in predetermined but any desired form. Some considerations and notes about network architecture and training process and applicability of currently developed pruning techniques and rule extraction algorithms are discussed. This technology, being developed by us for more than 10 years, allowed us to create dozens of knowledge-based expert systems.

The eight-year experience of application of this technology to problems in various fields allows us to estimate some propositions and draw come conclusions. The brief description is following.

First, neural network that can solve the problem with desired degree of accuracy has to be trained. Instead of use of mean square error as a estimation function it's better to use modified estimation functions which allow to control accuracy of decision in order to reach more simple resulting structure of the network at the second step of proposed technology. Also, in some cases it would be better to use multilayer neural networks, not only three-layered, in order to achieve more hierarchical set of rules with more simple rules at each level of hierarchy. The successful training of the network creates a hidden complex set of decision rules - the implicit knowledge.

Second, it's necessary to remove superfluous elements and inputs from the network. The pruning should lead to more simple interpretation of hidden rules, therefore it's not enough to remove only neurons or connections and it's necessary to introduce a methodology of complex pruning. So, we present the following set of available pruning operations: removing of inputs, neurons, synapses, biases, uniform simplification of a network (when the maximum number of synapses connected to neuron is decreased over the network). The user can establish the execution order of these operations itself. Then reduction of synaptic weights to values from a finite set of fixed values should be done. All these pruning operations are based on sensitivity analysis, only first-order derivatives are used [1]. Pruning is carried out by consecutive removal of inputs or elements and fails when it's impossible to reach zero of estimation function by retraining. The last step is replacement of sigmoid nonlinear transfer functions by threshold or piecewise-linear functions. Now the hidden implicit knowledge is refined and simplified, so it's possible to understand the meaning and generate explicit knowledge.

The third step is to write down the explicit knowledge in natural language. We propose to carry out such process by consecutive analysis of network structure manually. Because the meaning of input signals is already known, it's possible to substantially name the output signals of the first layer neurons, then second layer neurons and so on. The introduced requirement of network's uniform simplicity is proposed only for simplification of analysis phase, because the less number of input synapses of the neuron the easier to interpret and name its output signal. Also it's possible to extract rules from the network not manually, but automatically in fuzzy form of If-Then form, or in another form. Here we present a complete set of available types of rules that can be extracted from the network. The rule of a certain type may be interpreted in a different ways as a fuzzy, probabilistic or logical statement, but such interpretation involves expert knowledge about problem area and should be done manually by user.

*It should be noted that several networks of the same initial architecture and for the same data often lead to different decision rules (different explicit knowledge). It is not a lack.*

A particular case of rarefication of neural networks is selection of minimal sets of input signals sufficient for solving a problem. Such sets are not unique, as any knowledge based on data is not, though – other variants always are possible (except for really trivial and uninteresting cases). **Minimization of description** is one more sphere of services rendered by neuroinformatics.

One most important feature makes similar the expansion of neuroinformatics and the information revolution performed by personal computers – transition ***from elite marks to profane techniques*** (to the techniques which can be mastered by every commonly educated person, i.e. profane but not professional). Profanation not always is bad. Sacral oriental martial arts are being transformed into a quite profane occidental sport – and that is all right. Machinery evolves in the direction of increasing profanity, and that is normal.

Numerous population of diverse neural net algorithms inhabits the natural habitat of data processing, substituting the statistical provability [10] with neural net verisimilitude. That is sufficient in many fields, especially for preliminary conclusions and advices. Alchemistry of financial decisions is the best example. But also inverse process is necessary – from verisimilitude to provability. As one of urgent problem which is vital for further advance of neuroinformatics I would indicated **the problem of provability of neurocomputations**. Regular tools of verification of trained neural networks with exact evaluation of reliability are needed. A new field of science, "theory of reliability of artificial neural networks" should be created, otherwise many grave applications will be inaccessible for us.

An important argument for substantiation of expansion of artificial neural networks is the result that "neural networks can do anything", that by means of neural networks it is possible to compute any function of many variables up to any accuracy [11]. Yet there is such a widespread confusion in this issue that I should dwell upon it. As a symbol of belief many authors use the famous Kolmogorov-Arnold theorem on representation of continuous functions of several variables as superposition of continuous functions of one variable [13, 14]. However, this theorem states the possibility of *exact* representation of a function of many variables by means of *a very special set of functions* of one variable. These functions are very exotic, in particular, they are nowhere differentiable. On the contrary, by means of neural networks *approximate* representations of function of many variables are constructed. At that, *any limitations on the functions of one variable are absent*, one function is sufficient if only it is nonlinear [12]. Corresponding theorems do not have any concern to the outstanding Kolmogorov-Arnold theorem, but they are a generalization of as much outstanding Stone theorem [15], which, in turn, generalizes the Weierstrass theorem on approximation of functions by polynomials. Let $X$ be a compact set, $C(X)$ be the space of continuous functions on $X$, $M \subset C(X)$ be linear subspace, elements of *M* separate points in *X* (for any *x, y* from *X* there exists such *f*∈*M* that *f(x)*≠*f(y)*) and 1∈*M*.

*Stone theorem*. If *M* is subalgebra in *C(X)* (i.e. for any *f, g*∈*M* *fg*∈*M*), then *M* is dense in *C(X)*.

*Generalized approximation theorem.* If there exists continuous nonlinear function of one variable φ, for which under any *f*∈*M* also φ(*f*)∈*M*, then *M* is dense in *C(X)*.

*Corollary*. For any continuous nonlinear activation function of standard neuron (fig. 1) φ by means of neural networks it is possible to approximate uniformly any continuous function of many variables up to any accuracy on any closed bounded set.

There exist simple regular procedure of obtaining neural net approximation of continuous functions on bounded sets (as well as on finite ones – it is important for solution of problems of regression). That is training of so called $1^1/_2$-layer approximator with addition of neurons one-by-one. In this network there are only one layer of standard neurons and one output summator receiving all the output signals of neurons. Each new (n+1)-th neuron is trained to minimize mean square error of approximation of the function by the network (under fixed parameters of the first n neurons). The process converges: average square of the error tends to zero.

**Are there mathematical achievements generated by neuroinformatics?** Positive answer to this question would be a proof of maturity of this field of science. I believe that the answer is "yes". A result has matured inside neuroinformatics, which should come into courses of mathematical analysis and, therefore, is more important than most of technical inventions. Its importance for computational mathematics and all that what is called "computer sciences" exceeds the limits of neuroinformatics. I mean the basis of adaptation capabilities of neural networks – duality and fast differentiation of the functions computed by networks. Now and then, the method of back propagation of error is considered as a rule of training. That is not the most productive point of view: of course, motion against gradient decreases the value of function, but nevertheless it is natural to consider computation of derivatives apart from optimization. The method of back propagation of error is a way of computation of the gradient of estimate. Initially it was formulated for mean square deviations from the right answer, but it is easy to regard it as a method of computation of gradient of any function computable by network.

In order to clarify this issue, let discuss one "obvious" dogma, without destroying of which the effective training of networks would be impossible. Let the computational expenses (evaluated by the time spent by an universal computational device) for computation of one value of function of *n* variables $H(x_1,...,x_n)$ approximately equal to *T*. How much time it will take for the same device to compute grad*H* (if the program is rationally composed)? Most of mathematicians with university diploma would answer:
$$T_{gradH} \sim nT_H.$$

Incorrect! The right answer is:
$$T_{gradH} \sim CT_H,$$
where *C* is a constant which does not depend on the dimensionality *n* (in most cases *C*~3).

For all functions of many variables occurring in practice, including those computable by neural networks, necessary computational expenses on finding gradient are only two-three times larger than the expenses on computation of one value of function. That is astonishing, because the coordinates of the vector of gradient are *n* partial derivatives, and expenses on computation of one such derivative are in general case approximately the same as on computation of the value of function. Why it is less expensive to compute them together than separately?

The "miracle" is explained simply enough: it is necessary to organize computation of derivatives of function of many variables efficiently, avoiding duplication. For that, it is necessary to represent the computation of function in more details, in order to deal later not with a "black box" transforming the vector of arguments into the value of function, but with the graph of computations described in detail.

It is convenient to represent the search of grad*H* as a dual process over the structure of computation of *H*. Intermediate results coming from computation of gradient are nothing else than Lagrangian multipliers. It appears that if we represent *H* as composite function being a superposition of functions of small number of variables (and we are unable to compute functions of many variables in other way) and then accurately use the rule of differentiation of composite function, avoiding unnecessary computations and keeping useful intermediate results, then computation of the totality $\partial H/\partial x_i$ ($i=1,...,\underline{n}$) is a little bit complicated than computation of one of these functions – they all are constructed from identical blocks.

I do not know who was the first to think all this out. In neuroinformatics, the disputes on the priority are lasting so far. An end to re-discoveries was put by two works in 1986: Rumelhart, Hinton and Williams [16] and Bartsev and Okhonin [17]. However, the first publications belong to 70-th and even 60-th of our century. In the opinion of V.A. Okhonin, Lagrange and Legendre also have right to pretend to authorship of the method. Detailed presentation of mathematical content of these methods is given in chapter 3 of the book [18].

In order to represent the field of action in modern neuroinformatics as a whole, it is useful to draw a bird's-eye view of the main directions of investigations. A variant of such a map is given in fig. 4. The largest wave of works is devoted to **"neuronization"** – construction and application of neural net methods for solution of diverse problems amenable to the methods, but the other directions are of importance as well. In **user neural net programs** in the form of complete services neural network methods of solution of problems are realized, **modelling of real neural networks** is an inexhaustible source of ideas and analogies, and successful **new models of neurons and networks** though rarely appear but can have strong influence upon all other directions. As regards **realization of neural networks** either on usual parallel computing structures or in the form of specialized devices and neurocomputers, the future, undoubtedly, belongs to them. It should be noted that when constructing neurocomputers the main complication is not the realization of neural networks – that is simple, but representation of the processes of preconditioning of data in more or less standard form and their interpretation, formalization of standard formulations of problems and methods of training e.c. [19].

In conclusion, let me try to formulate what gives neural network realization of solution of a problem:
    1. Simple parallel realization on any parallel structures.
    2. Realization on special, fast and inexpensive neuro(co)processors.
    3. Realization on reliable systems from unreliable (including analogous) elements.
    4. Simple completion of adaptation blocks.
    5. Possibility of very fast realizations.

6. Possibility of "explicitation" of implicit knowledge by means of rarefiable neural networks.

7. Minimization of description.

..............

What else give neural networks?

..............

N+1. Universality – standard methods for solution of many problems, including non-standard ones. ...

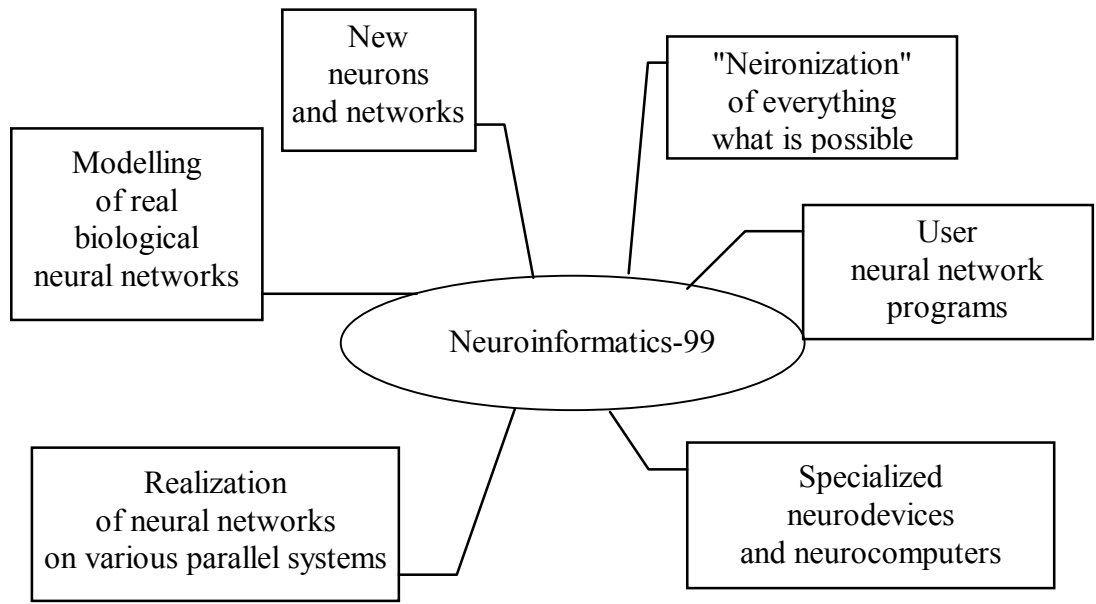

*Fig. 4. Neuroinformatics-99. Main directions of works.*

**References**[2]

---

[2] My purpose, on no account, was not to give a review of literature. The proposed list contains only the references which are essential for understanding of the note.